\documentclass[magazine,10pt,twocolume]{IEEEtran}
\usepackage{lettrine}
\usepackage{multirow}
\usepackage{graphicx}
\usepackage{subfigure}
\usepackage{amsmath}
\usepackage{amssymb}
\usepackage{amsfonts}
\usepackage{algorithm} 
\usepackage{algorithmic} 
\usepackage{cite}
\usepackage{xcolor}
\usepackage{bm}
\usepackage{float}
\usepackage{stfloats}
\usepackage{tabularx,array,booktabs}
\usepackage{makecell}

\makeatletter

\newcommand{\Rmnum}[1]{\expandafter\@slowromancap\romannumeral #1@}
\makeatother

\begin{document}

\title{\huge Multi-Beam Integrated Sensing and Communication: State-of-the-Art, Challenges and Opportunities\\
}
\author{Yinxiao Zhuo, Tianqi Mao, \emph{Member, IEEE}, Haojin Li, Chen Sun, \emph{Senior Member, IEEE} \\ Zhaocheng Wang, \emph{Fellow, IEEE}, Zhu Han, \emph{Fellow IEEE}, and Sheng Chen, \emph{Life Fellow, IEEE}
\thanks{This work was supported by the National Natural Science Foundation of China under Grant U22B2057. (\emph{Corresponding author: Zhaocheng Wang.})} %
\thanks{Y. Zhuo, and Z. Wang are with the Department of Electronic Engineering, Tsinghua University, Beijing 100084, China (E-mails: zhuoyx20@mails.tsinghua.edu.cn, zcwang@tsinghua.edu.cn). T. Mao is with Advanced Research Institute of Multidisciplinary Sciences, Beijing Institute of Technology, Beijing 100081, China (E-mails: maotq@buaa.edu.cn). H. Li and C. Sun are with the Sony R\&D Center, Beijing 100027, China (E-mails: Haojin.Li@sony.com, Chen.Sun@sony.com). Z. Han is with the Department of Electrical and Computer Engineering, University of Houston, Houston, TX 77004, USA (E-mail: hanzhu22@gmail.com). S. Chen is with the School of Electronics and Computer Science, University of Southampton, Southampton SO17 1BJ, U.K. (E-mail: sqc@ecs.soton.ac.uk).} %
\vspace{-10mm}
}
\maketitle

\begin{abstract}
Integrated sensing and communication (ISAC) has been envisioned as a critical enabling technology for the next-generation wireless communication, which can realize location/motion detection of surroundings with communication devices. This additional sensing capability leads to a substantial network quality gain and expansion of the service scenarios. As the system evolves to millimeter wave (mmWave) and above, ISAC can realize simultaneous communications and sensing of the ultra-high throughput level and radar resolution with compact design, which relies on directional beamforming against the path loss. With the multi-beam technology, the dual functions of ISAC can be seamlessly incorporated at the beamspace level by unleashing the potential of joint beamforming. To this end, this article investigates the key technologies for multi-beam ISAC system. We begin with an overview of the current state-of-the-art solutions in multi-beam ISAC. Subsequently, a detailed analysis of the advantages associated with the multi-beam ISAC is provided. Additionally, the key technologies for transmitter, channel and receiver of the multi-beam ISAC are introduced. Finally, we explore the challenges and opportunities presented by multi-beam ISAC, offering valuable insights into this emerging field.
\end{abstract}

\begin{IEEEkeywords}
Multi-beam integrated sensing and communication (ISAC), joint communication and radar (JCR), dual-functional radar and communication (DFRC), millimeter-wave (mmWave) communication, multi-beam beamforming, communication-sensing tradeoff.
\end{IEEEkeywords}

\IEEEpeerreviewmaketitle

\section{Introduction}\label{S1}

Integrated sensing and communication (ISAC), also known as joint communication and radar (JCR), dual-functional radar and communication (DFRC), has emerged as a promising technology for the sixth-generation (6G) network and beyond, which enables numerous cutting-edge applications, such as intelligent traffic, Internet of everything (IoE), and human-computer interaction \cite{background1, background2}. As an innovation concept, ISAC system is expected to combine the communication and sensing functions with shared hardware and frequency resources, thus reducing the expenses and improving the spectral efficiency. By integrating sensing function into the corn network, communication equipment has the potential of estimating the location and motion of surroundings, which can improve the quality of service (QoS) and expand future application scenarios. Moreover, the ISAC architecture enables cooperative information sharing between communication and sensing parts, which can be employed for additional performance gain of both subsystems.

To leverage ISAC technology, various ISAC frameworks have been proposed, and they can primarily be categorized into two categories: shared signaling scheme and separated signaling scheme. The shared signaling scheme aims to concurrently fulfill communication and sensing functions using a shared waveform, facilitating deep integration and optimal resource utilization \cite{shared-signal1}. However, this dual functionality presents significant challenges in designing shared signals, particularly in addressing mismatches between communication and sensing requirements \cite{shared-signal2}. Conversely, the separated signaling scheme prioritizes achieving greater independence and flexibility, albeit at the expense of resource efficiency. Within this framework, communication and sensing modules typically utilize distinct resources, enabling independent signal adjustments to meet their respective requirements.

This article focuses on multi-beam ISAC, which is a promising scheme of separated signaling ISAC. Inspired by millimeter wave (mmWave) space division multiplexing (SDM) technology, multi-beam ISAC is proposed to integrate the communication and sensing modules at the beamspace level, where the communication and sensing signals are carried by separated beams towards the user terminals and target detection, respectively. However, in the typical ISAC application scenario, the requirements of communication and sensing beams are often distinct in terms of beamwidth, coverage range, and stability, which impose constraints on each other and necessitates joint beamforming design. Furthermore, although the SDM design provides a certain degree of isolation between communication and sensing signals, the possible energy leakage and diffraction/reflection effects can still induce inevitable interactions between the dual functions.


\begin{figure*}[t]
\vspace*{-2mm}
\centering
\includegraphics[width=0.95\linewidth]{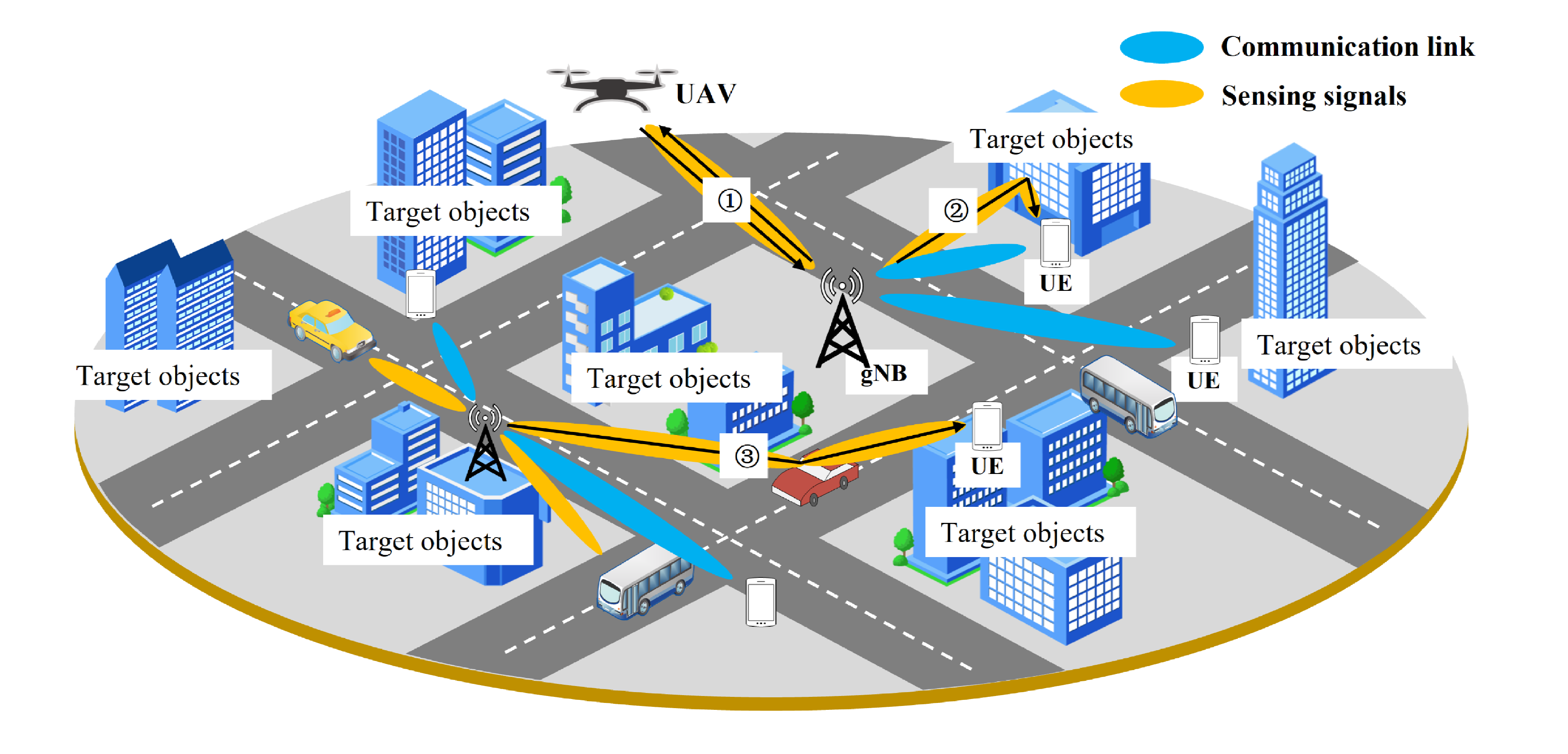}
\vspace{-6mm}
\caption{Multi-beam ISAC application scenarios with different configurations. {\small{\textcircled{\scriptsize{1}}}} Co-located Tx-RRx: RRx are deployed at Tx side which receives the echoes of its own sensing signals. {\small{\textcircled{\scriptsize{2}}}} Co-located CRx-RRx: The CRx and RRx are merged into one device which receives both communication and sensing signals. {\small{\textcircled{\scriptsize{3}}}} Non-co-located RRx: The remote RRx device receives only the reflected sensing signals.}
\label{fig:fig1} 
\vspace*{-2mm}
\end{figure*}

Compared to the existing surveys on ISAC \cite{background1, background2,shared-signal1,shared-signal2,survey2}, this article innovatively analyzes multi-beam ISAC from the perspective of ISAC transceivers, highlighting configuration differences and the required technologies. Specifically, this article presents a comprehensive review of the current state-of-the-art developments in the field of multi-beam ISAC. We systematically analyze its strengths and limitations, drawing comparisons with other established ISAC approaches. Subsequently, based on transceiver configurations, beamforming design, correlative channel modeling method and inter-beam interference management are elucidated as key technologies to underpin the three fundamental elements of an ISAC system, the transmitter, channel, and receiver, respectively. Finally, future challenges and opportunities are highlighted for further exploration, including transceiver optimization design, ISAC cooperation strategies and high-frequency effect handling.

\section{Multi-Beam ISAC}\label{S2}

\subsection{ISAC System Model}\label{S2.1}

In this article, we concentrate on a typical ISAC service that requires high data rate communication and high precision parameter estimation of passive targets, where channel capacity and Cramer-Rao lower bound (CRLB) can be used for respective performance evaluation. The ISAC systems are generally composed of a shared transmitter (Tx), a communication receiver (CRx) and a sensing receiver (RRx). The shared Tx simultaneously transmits the communication and sensing signals, while the CRx and RRx receive their signals of interest, respectively \cite{survey2}. Based on the co-location relationship of the three transceivers, the ISAC system models can be split into the following three categories: 
\begin{itemize}
\item \textbf{Co-Located Tx-RRx:} In this scheme, the RRx is deployed at the Tx side. As shown in Fig.~\ref{fig:fig1}{\normalsize{\textcircled{\footnotesize{1}}}}, the sensing signals illuminate the target and are reflected back to Tx, and the RRx at Tx side extracts the sensing parameters of the target from the echo signals. From a sensing perspective, Tx can be considered as a monostatic radar.

\item \textbf{Co-Located CRx-RRx:} In this scheme, the CRx and RRx are merged into a dual-function Rx which receives both the communication signals and the reflected sensing signals. Illustrated in Fig.~\ref{fig:fig1}{\normalsize{\textcircled{\footnotesize{2}}}}, a shared link is established between the gNodeB (gNB) and user equipment (UE) for data transmission and bistatic radar detection.

\item \textbf{Non-Co-Located RRx:} In non-co-located RRx configuration, Tx, CRx and RRx are all different devices, and Tx should establish two different links with CRx and RRx, respectively. Fig.~\ref{fig:fig1}{\normalsize{\textcircled{\footnotesize{3}}}} shows an example of non-co-located RRx, where the remote RRx receives only the reflected sensing signals without a communication link.
\end{itemize}

It should be pointed out that Fig.~\ref{fig:fig1} takes downlink transmission as an example. For the uplink scenario, UE serves as Tx to transmit the communication and sensing signals. 

The various RRx configuration implies diverse application scenarios as well as different signal models. 
For example, roadside gNB in Internet of vehicle (IoV) is an example for co-located Tx-RRx, which possesses knowledge of both the transmitted communication signals and sensing signals, and actively obtains sensing information for scheduling and control. On the other hand, the co-located CRx-RRx receives superimposed communication-sensing signals and attempts to distinguish them apart in high signal-to-noise ratio (SNR) regions. This configuration makes it possible for UE to change the communication topology timely with sensing results. Compared to non-co-located RRx, these co-located configuration enables information fusion and joint signal processing. These differences result in significant variations in signal processing methods, waveform designs and cooperation approaches between the communication and sensing modules.

\subsection{Multi-Beam ISAC}\label{S2.2}

Thanks to the short wavelength and abundant spectrum resources, mmWave signals are especially suitable for high-resolution ISAC applications. To fully utilize the beamforming capability of the large antenna array at mmWave frequencies, Zhang \emph{et al.} \cite{multi-beam-ISAC1} proposed a multi-beam ISAC scheme for the first time, which generates a beamforming pattern with multiple mainlobes for both communication and sensing. Within the coherent time of the communication channel, the multiple beams can be decomposed into two parts: time-invariant communication beams with high and stable beamforming gain, and other time-varying sensing beams for tracking the moving target at the wide sensing field of view (FoV).

Obviously, multi-beam ISAC separates the communication and sensing functions in the spatial domain, which enables simultaneous communications and sensing with reduced interference level and sufficient flexibility regarding beam assignment. This reduces interference level and allows for efficient reuse of time-frequency resources, which significantly enhances spectrum efficiency. Furthermore, the flexible beam assignment enables relatively autonomous beamforming for communication and sensing, making it suitable for diverse applications with varying communication data rates, latency requirements, sensing duration and resolution needs.

From the perspective of integration technology, multi-beam ISAC has progressed beyond the stage of simple coexistence and entered the cooperation stage. Although the communication and sensing modules still compete for power allocation, the joint design technology allows for power sharing between communication and sensing, which improves power efficiency. For instance, by joint beamforming design, the communication signals can be reused for sensing, and the sidelobe power leakage from the sensing beam can be harnessed to enhance communication SNR. Researchers have made significant progress on the multi-beam ISAC technology. Specifically, the study \cite{multi-beam-ISAC1} adopted the principle of `coherent combining phase' for beamforming, which adjusts the phases of the sensing beam to make its sidelobe coherent to the communication counterpart. This approach allows to utilize a portion of sensing energy to enhance the beamforming gain for data transmission. 
The work \cite{multi-beam-ISAC3} proposed a joint beamforming multi-beam ISAC scheme that allows communication signals to be reused for sensing, and minimizes the sensing normalized mean square error (NMSE) at the expense of a slight reduction in the communication rate. The study \cite{multi-beam-ISAC4} employed an alternating optimization algorithm to simultaneously decode communication signals and estimate the cross-interference channel, aiding the CRx in mitigating interference effects. These innovative designs mitigate the mutual constraints between communication and sensing, and foster collaboration between them.

\subsection{Comparison among Different ISAC Schemes}\label{S2.3}

To facilitate practical ISAC applications, in addition to the aforementioned multi-beam ISAC schemes, researchers from academia and industry have also proposed numerous candidate schemes, which mainly include resource division \cite{TD-ISAC} and waveform sharing schemes \cite{C-Centric-ISAC, shared-signal1}. Resource division schemes allocate different time-frequency resource elements for communication and sensing, respectively \cite{TD-ISAC}. Therefore, these schemes ensure minimal cross-interference and are typically compatible with the fifth-generation (5G) orthogonal frequency division multiplexing (OFDM) system. However, due to the competition for time-frequency resources between communication and sensing, these schemes exhibit relatively poor spectrum and energy efficiency. On the other hand, waveform embedding schemes utilize existing communication or radar waveforms as their foundations and incorporate the other function into them. The scheme \cite{C-Centric-ISAC} forces the radar to utilize communication OFDM signals for sensing, and the work \cite{shared-signal1} achieves target sensing through backscatter from communication signals. With shared waveform, these schemes are cross-interference-free and enable both communication and sensing modules to simultaneously occupy the entire bandwidth. Nevertheless, the flexibility is degraded due to the constraint of the basic waveform, which makes it difficult for communication and sensing to independently adjust signal strength, duration, coverage range, and other parameters.

\begin{figure*}[!t]
\vspace*{-1mm}
\centering
\includegraphics[width=0.86\linewidth]{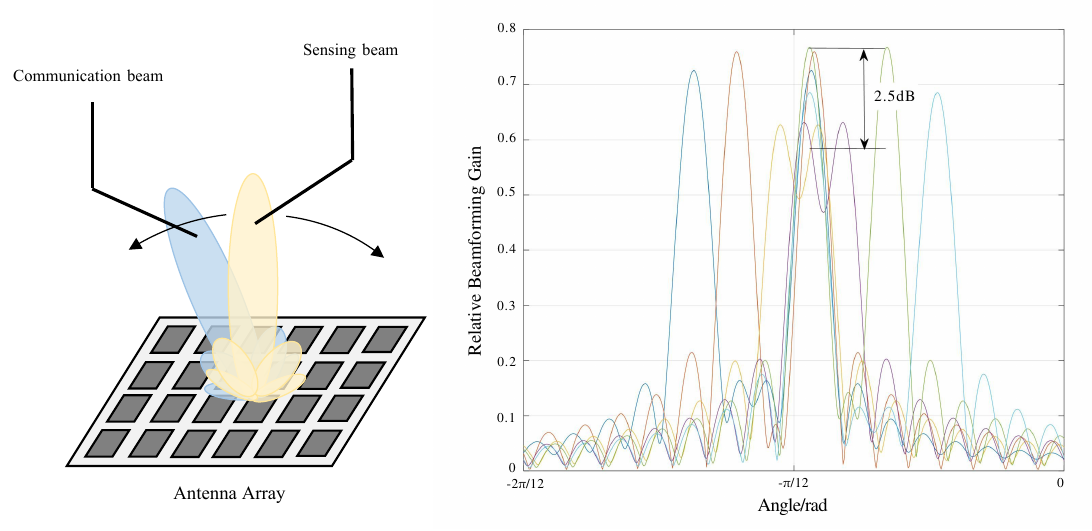}
\vspace{-5mm}
\caption{Illustration of beamforming gain fluctuations caused by the time-variant sensing beam.}
\label{fig:fig2} 
\vspace*{-3mm}
\end{figure*}

Compared with these ISAC counterparts, multi-beam ISAC presents its superiority in terms of both spectrum efficiency and flexibility. Multi-beam ISAC spatially separates communication and sensing, ensuring low cross-interference levels while efficiently utilizing time-frequency resources to enhance spectrum efficiency. Additionally, since the communication and sensing subsystems are relatively independent, they can independently and flexibly adjust transmission parameters, which makes this architecture capable of meeting various QoS requirements. Moreover, the precoding architecture and signal processing procedures for communication and sensing are similar and decoupled, which allows the multi-beam ISAC system to seamlessly inherit the existing communication and sensing technologies.

Despite its attractive merits, however, the practical application of multi-beam ISAC still encounters several challenges. Properly handling the cross-effect between communication and sensing beams is crucial to fully unleash the potential of multi-beam ISAC. We elucidate the challenges from the three essential components of the ISAC systems, namely, transmitter, channel and receiver, in the sequel.
\begin{itemize}
\item \textbf{Transmitter:} The mismatch in the requirements between communication and sensing poses challenges for signal transmitter design. For communication, signals require stable and high-gain beams to achieve high data rate and low latency. In contrast, the sensing beam usually changes rapidly to sweep a wide range of interests or seamlessly capture the high-mobility target. This mismatch complicates the design and scheduling of transmit beams.

\item \textbf{Channel:} Since communication and sensing signals share the same propagation environment, their respective channels should inherently possess correlated parameters, which makes it possible to efficiently and accurately estimate both the communication and sensing channels. Therefore, the corresponding channel estimation algorithm capable of accurately modeling of the channel correlation is a crucial challenge for improving the performance of multi-beam ISAC.

\item \textbf{Receiver:} Although multi-beam ISAC reduces cross-interference through beam division design, the sidelobe power leakage and signal reflection inevitably cause cross-interference among themselves, which poses challenges for interference mitigation techniques.
\end{itemize}

\begin{figure*}[!b]
\vspace*{-4mm}
\centering
\includegraphics[width=0.86\linewidth]{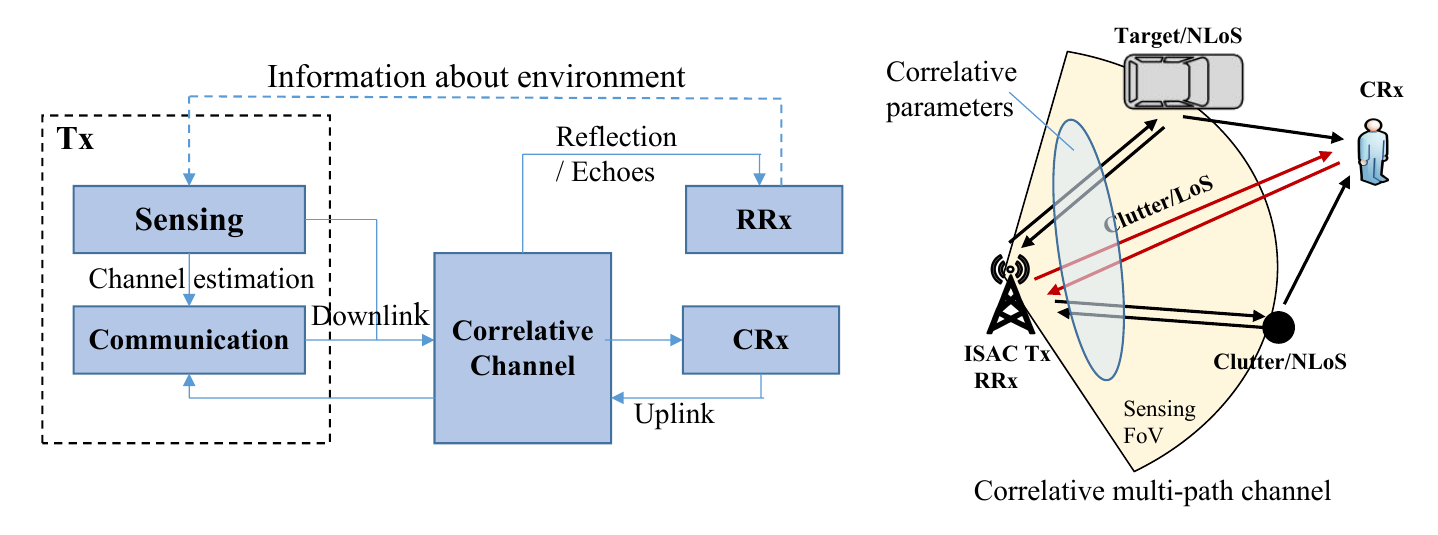}
\vspace{-3mm}
\caption{Multi-beam ISAC system model under correlative channel, where both communication and sensing channels share correlative parameters.}
\label{fig:fig3} 
\vspace*{-1mm}
\end{figure*}

\section{Key Technologies of Multi-Beam ISAC}\label{S3}

To address these challenges, researchers have proposed various new technologies to enhance the performance of multi-beam ISAC. In this section, we place an emphasis on beamformer design, channel modeling, and interference management for multi-beam ISAC, which are especially crucial to address the challenges of the three essential components and fully unleash the potential of beamforming technology.

\subsection{Beamforming Design}\label{S3.1}

To address the mismatch between communication and sensing beams, one straightforward approach would be to generate communication and sensing beams separately using different RF chains and then superimpose them to form the multi-beam pattern. However, the fast variation of the sensing beam results in highly unstable beamforming gain for the communication direction. As illustrated in Fig.~\ref{fig:fig2}, when communication and sensing beams having the same power level are considered, due to the influence of the sidelobes from the sensing beam, the beamforming gain in the communication direction experiences fluctuations of up to approximately 2.5\,dB, which introduces significant challenges in communication channel measurement and channel equalization. 

To mitigate these fluctuations, a proposed solution, as outlined in \cite{multi-beam-ISAC1}, involves initially designing the desired superimposed beam pattern and then approximating this pattern with hybrid beamforming architecture. This joint beamforming method effectively suppresses sidelobes and stabilizes the beamforming gain. Furthermore, the research presented in \cite{multi-beam-ISAC3} suggests that it is not necessary to fully design the entire superimposed beam pattern. Instead, it suffices to adhere to the communication and sensing requirements concerning the beam gains in specific directions or the cumulative power within a designated range. By relaxing the constraints on the superimposed beam pattern, the approximation error can be reduced accordingly.

\begin{figure*}[!ht]\setcounter{figure}{3}
	\vspace*{-1mm}
	\centering
	\includegraphics[width=0.9\linewidth]{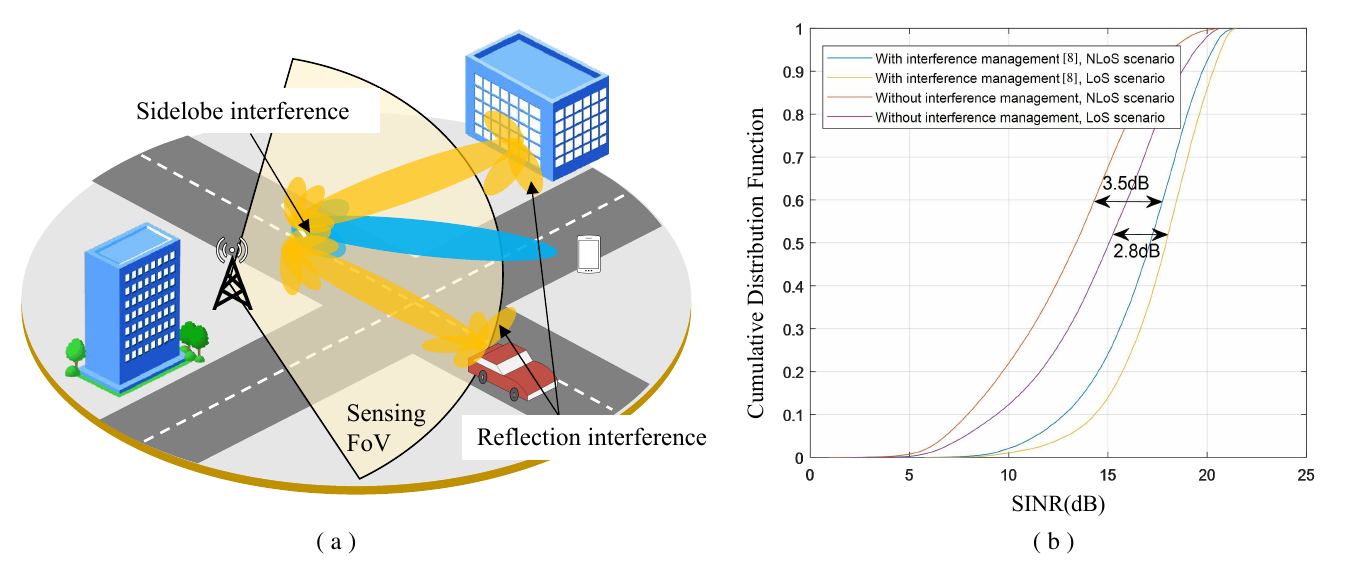}
	\vspace{-4mm}
	\caption{An example of inter-beam interference in multi-beam ISAC system. The sidelobe leakage and reflection of the sensing beams cause inter-beam interference. Interference management methodology improves SINR at CRx in both LoS and NLoS scenarios\cite{multi-beam-ISAC4}. }
	\label{fig:fig4} 
	\vspace*{-2mm}
\end{figure*}

\subsection{Correlative Channel Modeling}\label{S3.2}

For simplicity, many researchers have employed independent line-of-sight (LoS)-dominated channel models for both communication and sensing tasks. In reality, however, since communication and sensing signals share the same propagation environment, their respective channels inherently possess correlated parameters. Taking the 3rd Generation Partnership Project (3GPP) clustered delay line (CDL) channel model as an example, it becomes evident that the azimuth/zenith angles and latency of multi-paths are influenced by the surrounding scatterers. Consequently, the CDL channels for both communication and sensing exhibit correlated angle and delay parameters.

Fig.~\ref{fig:fig3} provides a diagram of the multi-beam ISAC system under the correlative channel model. Compared to traditional communication systems, multi-beam ISAC systems have an additional sensing link sharing the correlative channel. The reflection signals received by RRx obtain information about the environment, which can in turn benefit channel estimation. To explore improvement methods for the correlated channel model, the study \cite{Fundamental} proposes an information-theoretic model for the point-to-point memoryless ISAC channel. The papers \cite{theory1} and \cite{theory2} further formulate the ISAC information theoretic model under difference channel assumptions, and then present a capacity-distortion tradeoff for the ISAC channel, asserting that the joint signal transmission design in ISAC offers superior performance compared to a separate communication-sensing approach. Additionally, the work \cite{Correlative-Channel} presents a specific joint design approach. Under the assumption that the communication and sensing multipath channels share the same reflection paths, the authors of \cite{Correlative-Channel} propose a sensing-assisted beam alignment scheme to enhance communication performance. The analysis of the CRLB and simulation results demonstrate the superiority of this sensing-assisted communication scheme.

\subsection{Inter-Beam Interference Management}\label{S3.3}

For various RRx configurations, the impact and mitigation strategies of cross-interference exhibit nuanced differences driven by variations in the signal models. The co-located Tx-RRx is the most widely considered configuration, where both the communication and sensing signals are known to the RRx, which facilitates interference-free sensing operations. However, this favorable scenario for sensing introduces sensing interference on the CRx. As is illustrated in Fig.~\ref{fig:fig4}(a), the sidelobe and reflection interference are received by CRx, which significantly degrades the signal-to-interference and noise ratio (SINR) at CRx. Therefore, the implementation of interference mitigation algorithms on the CRx side becomes imperative. The work \cite{CS-Interference-Cancellation} leverages the angular sparsity of the ISAC channel and employs compress sensing (CS) techniques to jointly estimate sensing interference and demodulate communication signals. Furthermore, the study \cite{multi-beam-ISAC4} assumes that CRx has prior information about the sensing signal sequences, and utilizes alternative optimization to iteratively estimate the interference and demodulate the communication signals, thereby alleviating interference and enhancing communication performance. Observed from Fig.~\ref{fig:fig4}(b), the interference management \cite{multi-beam-ISAC4} improves SINR at CRx by around 2.8dB in LoS scenario and 3.5dB in Non-LoS (NLoS) scenario.

\begin{figure*}[!ht]\setcounter{figure}{4}
	\vspace{-2mm}
	\centering
	\includegraphics[width=0.95\linewidth]{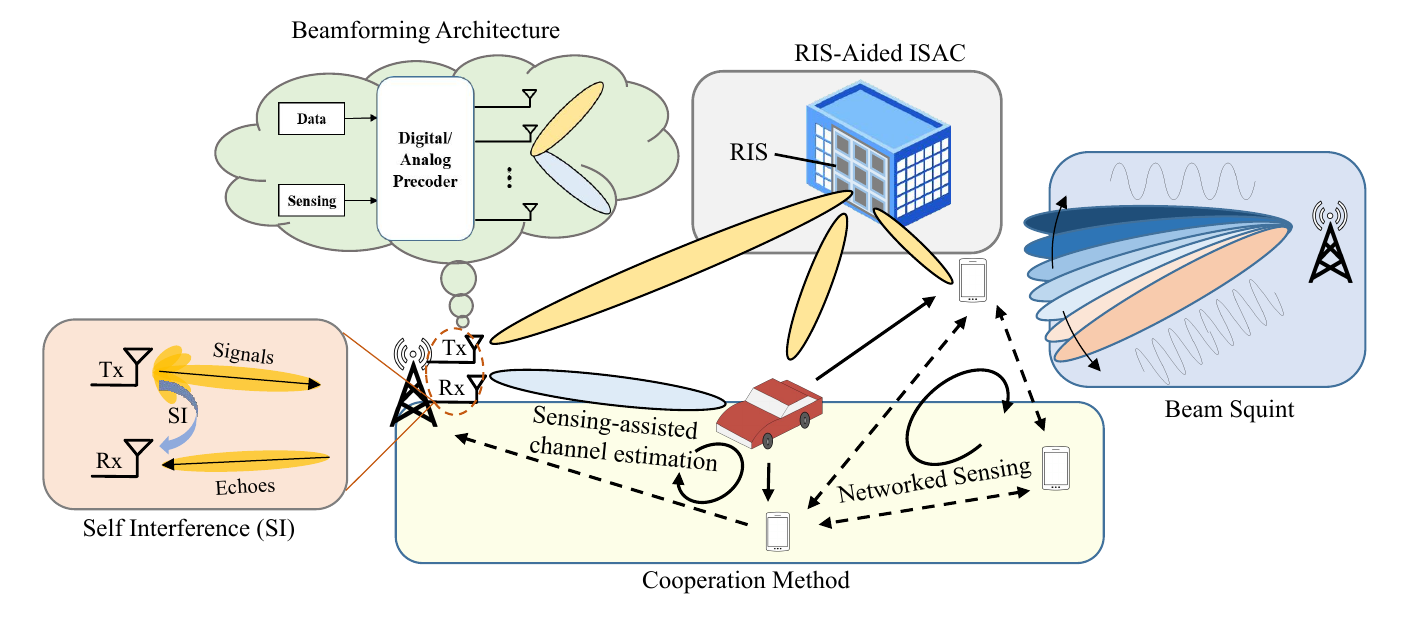}
	\vspace{-2mm}
	\caption{Open issues and challenges for multi-beam ISAC scheme.}
	\label{fig:fig5} 
	\vspace{-2mm}
\end{figure*}

\section{Open Issues and Challenges}\label{S4}

Despite of its promising breakthroughs in recent years, the multi-beam ISAC is still facing a plethora of open issues that are worth further investigation. To improve communication data rate and sensing precision, it is important for the ISAC transceivers to handle the cross-effect. From the aspects of transmitter, channel and receiver, below we select some important technical challenges point by point, which are also illustrated in Fig.~\ref{fig:fig5}.

\subsection{Beamforming Architecture}\label{S4.1}

Multi-beam ISAC systems require the simultaneous activation of time-variant sensing beams and time-invariant communication beams, leading to asynchronous control for distinct beam components. However, both the two primary methods for generating multi-beam patterns have their limitations. The first approach is to generate separate communication and sensing beams, and subsequently superimpose them, which is straightforward and supports asynchronous beam scheduling. However, it poses challenges in controlling sidelobe leakage, which results in cross-interference between communication and sensing beams. Conversely, the second approach approximates the superimposed multi-beam pattern directly. This technique provides better control over beamforming gains in both communication and sensing directions, effectively suppressing sidelobe effects. Nonetheless, it introduces approximation errors and demands significant additional overhead due to asynchronous beam scheduling. Additionally, the power imbalances between communication and sensing beams poses more challenges for power amplifiers and signal processing, especially for Co-located CRx-RRx configuration. To achieve a better tradeoff between beam pattern control and scheduling overhead, there is a necessity to explore advanced beamforming architectures and beamforming techniques for the specific requirements of multi-beam ISAC.

\subsection{Self-Interference Management}\label{S4.2}

Since multi-beam ISAC transmitter operates continuously, continues-wave radar is preferred to be applied in ISAC systems with the co-located Tx-RRx configuration, which however introduces full-duplex as an inevitable problem. Specifically, as the ISAC transmitter operates continuously, the radar receiver may suffer from pronounced self-interference resulting from the leakage of transmitted signals. This can undermine the weaker echoes from relatively distant targets, causing detection performance degradation. To tackle this issue, a possible approach is to involve hardware isolation between the transmitter and receiver, ensuring that the transmitted signal is effectively isolated from the receiver front-end to minimize self-interference. Alternatively, advanced signal processing techniques and self-interference cancellation algorithms are helpful in distinguishing weak echoes from superimposed signals, which can significantly enhance the practical applicability of multi-beam ISAC systems.

\subsection{Reconfigurable Intelligent Surface Aided Multi-Beam ISAC}\label{S4.25}

Integrating a reconfigurable intelligent surface (RIS) into ISAC systems provides controllable reflection paths to supplement coverage in blind zone, which is particularly helpful in addressing the blockage problem \cite{shared-signal2}. With its programmable meta-surface, the RIS can capture received energy and redirect it through reoriented beams, providing extra degrees of freedom (DoF). Moreover, the active RIS can provide higher gains to compensate for multiplicative fading. However, the deployment of RIS also complicates the ISAC system, which necessitates more intricate beam scheduling to fully harness the potential of the multi-beam ISAC. Specifically, the measurement and control methodology for RIS is essential to strike a balance between reflection gain and signal delay/computational complexity. Furthermore, the optimal placement, orientation and density of RIS element must be carefully designed to maximize benefits while adhering to spatial and cost constraints. Therefore, conducting research on RIS-aided multi-beam ISAC holds significant promise for a more efficient, cost-effective, and versatile ISAC solution.

\subsection{Cooperation Method}\label{S4.3}

The multi-beam ISAC architecture provides the potential for cooperation between communication and sensing. However, the current cooperation methods are still in their early stages. To fully leverage the gain of cooperation and approach the information-theoretic bounds presented in \cite{Fundamental}, a more seamless and intensive cooperation method is highly necessary. For communication, sensing-assisted channel estimation is a feasible approach, which can reduce channel estimation overhead while enhancing estimation accuracy to achieve higher CRx SNR. On the sensing side, communication can facilitate the networking of sensing devices, enabling multi-node collaborative sensing, which allows a single sensing device to have a broader coverage area and higher sensing accuracy.

\subsection{Terahertz Beam-Squint Effect}\label{S4.4}

The Terahertz-band (THz-band) ISAC system holds great promise for future applications. On one hand, the wide bandwidth and short wavelength have the potential to simultaneously support ultra-high data rate communication and ultra-high-resolution imaging. On the other hand, the ability to penetrate non-metallic materials, such as clothing and carton, extends sensing application scenarios. However, the high frequency and wide bandwidth also induce new challenges compared with classical microwave frequencies, particularly the issue of beam-squint. Specifically, the wide bandwidth leads to frequency selectivity in the phase-shift-based discrete Fourier transform (DFT) beamforming, which means that the different subcarriers will have distinct beam directions. As a result, the deviation in beam direction can cause significant degradation on the array gains for different subcarriers, and lead to beam overlapping in the multi-beam scheme. One potential solution is to implement programmable time delays for time-phase precoding, aimed at suppressing the beam squint effect. Additionally, leveraging beam squint to activate different beams on different subcarriers provides an opportunity for multi-beam communication sensing systems. 

\section{Conclusions}\label{S5}

In this article, we have commenced by reviewing the development of the multi-beam ISAC system, which has attracted much attention due to its unique advantages. With superiority in spectrum efficiency, flexibility and compatibility, multi-beam ISAC has emerged as one of the most promising ISAC architectures for practical applications. We have illustrated the core technologies of multi-beam ISAC from the perspectives of the transmitter, channel, and receiver, providing insight into how to effectively manage the competition and cooperation between the communication and sensing modules. Finally, the challenges and future research directions for multi-beam ISAC are outlined.

\ifCLASSOPTIONcaptionsoff
  \newpage
\fi

\section*{Biographies}
\vspace{-20mm}
\begin{IEEEbiographynophoto}
	{Yinxiao Zhuo} received his B.S., M.S., degree from Tsinghua University in 2020, where he is currently working toward the Ph.D. degree with the Department of Electronic Engineering, Tsinghua University. His current research interests include mmWave communications and multi-beam integrated sensing and communication.
\end{IEEEbiographynophoto}
\vspace{-20mm}
\begin{IEEEbiographynophoto}	
	{Tianqi Mao} received the B.S., M.S., and Ph.D. degrees (Hons.) from Tsinghua University in 2015, 2018, and 2022, respectively. He is currently working with the Advanced Research Institute of Multidisciplinary Sciences, Beijing Institute of Technology, Beijing, China. His current research interests include modulation and signal processing for wireless communications, terahertz communications, nearspace wireless communications, and visible light communications. He was a recipient of the 8th Young Elite Scientists Sponsorship Program by the China Association for Science and Technology.
\end{IEEEbiographynophoto}
\vspace{-20mm}
\begin{IEEEbiographynophoto}
	{Haojin Li} received the M.S. degree in information and communication engineering from the University of China Academy of Telecommunication Technology, Beijing, China, in 2020. He joined Sony China Research Laboratory in 2020 as a 5G researcher and since then has been responsible for Sony global 5G System Level Simulator development, in June 2022, he was promoted to Deputy Principal Research and Development Engineer. His research interests include non-terrestrial networks and integrated sensing and communication.
\end{IEEEbiographynophoto}
\vspace{-20mm}
\begin{IEEEbiographynophoto}
	{Chen Sun} received the Ph.D. degree in electrical engineering from Nanyang Technological University, Singapore, in 2005. From August 2004 to May 2008, he was a researcher at ATR Wave Engineering Laboratories, Japan working on adaptive beamforming and direction finding algorithms of parasitic array antennas as well as theoretical analysis of cooperative wireless networks. In June 2008, he joined the National Institute of Information and Communications Technology (NICT), Japan, as expert researcher working on distributed sensing and dynamic spectrum access in TV white space. He is currently the deputy head of Beijing Lab at Sony R\&D Center.
\end{IEEEbiographynophoto}
\vspace{-20mm}
\begin{IEEEbiographynophoto}
	{Zhaocheng Wang} [F’21] received his B.S., M.S., and Ph.D. degrees from Tsinghua University in 1991, 1993, and 1996, respectively. From 1996 to 1997, he was a Post Doctoral Fellow with Nanyang	Technological University, Singapore. From 1997 to 2009, he was a Research Engineer/Senior Engineer with OKI Techno Centre Pte. Ltd., Singapore. From 1999 to 2009, he was a Senior Engineer/Principal Engineer with Sony Deutschland GmbH, Germany. Since 2009, he has been a Professor with Department of Electronic Engineering, Tsinghua University. He was a recipient of IEEE Scott Helt Memorial Award, IET Premium Award, IEEE ComSoc Asia-Pacific Outstanding Paper Award and IEEE ComSoc Leonard G. Abraham Prize. 
\end{IEEEbiographynophoto}
\vspace{-20mm}
\begin{IEEEbiographynophoto}
	{Zhu Han} [F’14] received the B.S. degree in electronic engineering from Tsinghua University, Beijing, China, in 1997, and the M.S. and Ph.D. degrees in electrical and computer engineering from the University of Maryland, College Park, MD, USA, in 1999 and 2003, respectively. From 2000 to 2002, he was a R\&D Engineer of JDSU, Germantown, MD, USA. From 2003 to 2006, he was a Research Associate at the University of Maryland. From 2006 to 2008, he was an Assistant Professor at Boise State University, Boise, ID, USA. He is currently a John and Rebecca Moores Professor with the Electrical and Computer Engineering Department and the Computer Science Department, University of Houston, Houston, TX, USA. He has been an American Association for the Advancement of Science (AAAS) Fellow and an Association for Computing Machinery (ACM) Distinguished Member since 2019.
\end{IEEEbiographynophoto}
\vspace{-20mm}
\begin{IEEEbiographynophoto}
	{Sheng Chen} [F’08] received his B.Eng. degree in control engineering from East China Petroleum Institute, China, in 1982, and his Ph.D. degree in control engineering from City University, U.K., in 1986, and the Doctor of Sciences (D.Sc.) degree from the University of Southampton, U.K., in 2005. From 1986 to 1999, He	held research and academic appointments with the Universities of Sheffield, Edinburgh and Portsmouth, all in U.K. Since 1999, he has been with the School of Electronics and Computer Science, University of Southampton, U.K., where he holds the post of a Professor in intelligent systems and signal processing. He is a Fellow of the Royal Academy of Engineering (FREng), of the Asia-Pacific Artificial Intelligence Association and of the IET.
\end{IEEEbiographynophoto}

\end{document}